\documentclass[apl,showpacs,twocolumn]{revtex4}
\usepackage{amssymb}
\usepackage{graphicx,color,ulem}
\begin{document}


\title{Suppression of superconductivity in FeSe films under tensile strain}

\author{Y. F. Nie}
\author{E. Brahimi}
\author{J. I. Budnick}
\author{W. A. Hines}
\author{M. Jain}
\author{B. O. Wells}
\affiliation{Department of Physics, University of Connecticut,
Storrs, CT 06269}

\date{\today}

\begin{abstract}
We have studied the effect of tensile strain on the
superconductivity in FeSe films. 50 nm, 100 nm, and 200 nm FeSe
films were grown on MgO, SrTiO$_3$, and LaAlO$_3$ substrates by
using a pulsed laser deposition technique. X-ray diffraction
analysis showed that the tetragonal phase is dominant in all of our
FeSe films. The 50 nm FeSe films on MgO and SrTiO$_3$ are under
tensile strain, while the 50 nm FeSe film on LaAlO$_3$ and the other
thick FeSe films are unstrained. Superconducting transitions have
been observed in unstrained FeSe films with T$_{onset}$ $\approx$ 8
K, which is close to the bulk value. However, no sign of
superconductivity has been observed in FeSe films under tensile
strain down to 5 K. This is evidence to show that tensile strain
suppresses superconductivity in FeSe films.
\end{abstract}

\maketitle


Since the discovery of superconductivity in doped LaOFeAs with a
T$_c$ of 26 K by  Kamihara et al.~\cite{Hosono01}, the pnictide
superconductors have attracted much
attention~\cite{Hosono02,Chen01,Dai01}. Just as the CuO$_2$ plane
plays a key role in high T$_c$ superconducting copper
oxides~\cite{Bednorz01}, the FeAs layer is believed to be crucial to
the superconductivity in iron-arsenide~\cite{Dai01}. Stimulated by
this discovery, other iron-based planar compounds have been
investigated in a search for
superconductivity~\cite{Rotter01,Pitcher01,Hsu01}.

Recently, superconductivity was observed in $\alpha$-FeSe with a
transition temperature T$_c$ $\approx$ 8 K~\cite{Hsu01}.
$\alpha$-FeSe with the tetragonal PbO structure has an Fe-based
planar sublattice equivalent to the layered iron-based quaternary
oxypnictides. Thus, it is expected that the study of
superconductivity in FeSe will be valuable in understanding the
superconducting mechanism of pnictide superconductors. FeSe has a
complicated phase diagram~\cite{Terzieff01,Schuster01}. The
superconductivity of FeSe can be enhanced by partially replacing Se
with Te, where T$_c$ reaches a maximum of 15.2 K at about 50\% Te
substitution~\cite{Yeh01}. Also, it has been reported that the T$_c$
of $\alpha$-FeSe has a strong dependence on hydrostatic pressure,
and can be enhanced up to 27 K by applying a pressure of 1.48
GPa~\cite{Mizuguchi01}. Another study reported that the application
of hydrostatic pressure first rapidly increases T$_c$ to a broad
maximum of 37 K at 7 GPa, before decreasing to 6 K at 14
GPa~\cite{Margadonna01}. A phase transition from tetragonal to
orthorhombic has been reported at 12 GPa, with the high pressure
orthorhombic phase having a higher T$_c$, reaching 34 K at 22 GPa
~\cite{Garbarino01}. There are additional studies of the pressure
effects for FeSe bulk materials ~\cite{Millican01}. All of the
studies imply that strain (compression) is a very important
parameter for iron-based superconductivity.

In all of these high-pressure studies, the materials are under
compression; however, how FeSe will behave under tension is still
unknown. It is very important to study the tensile strain effect in
order to understand the mechanism of iron-based superconductivity.
It is difficult to introduce large tensile strain in a bulk
material. On the contrary, it is relatively easy to introduce large
strain in epitaxial films, both compressive and tensile. Strain in
films differs from the hydrostatic pressure in bulk in that film
strain is biaxial. However, for a layered material, biaxial strain
may be the most interesting. The biaxial strain comes from the
energetically beneficial registry growth of the films on selected
substrates. Coherently strained films in which the film and
substrate in-plane lattice parameters are forced to match can be
maintained up to a critical thickness, which is of the order of a
few nanometers, depending on the amount of the mismatch. Above the
critical thickness, the strain energy becomes so large that it is
then energetically favorable to nucleate misfit
dislocations~\cite{Matthews01}. By properly choosing the substrate
and film thickness, we can grow films with various tensile strains.
In this report, we present a study of the tensile strain effect on
the superconductivity in FeSe films.


\begin{figure}[t]
\begin{center}
\includegraphics[width=80mm]{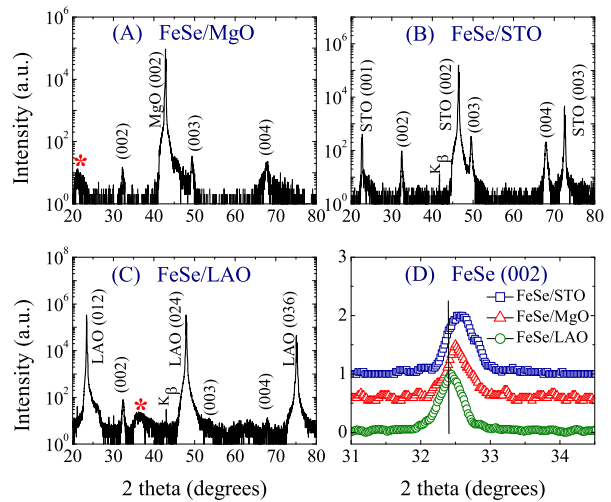}
\caption{\label{fig:fig1.eps}(color online) X-ray diffraction
profiles for 50 nm FeSe films on MgO, STO, and LAO substrates at
room temperature. The panel (D) illustrates the effect of in-plane
strain on the FeSe (002) peaks, with peaks normalized to an
intensity of one and offset for clarity.}
\end{center}
\end{figure}


The FeSe films studied were grown on (001) MgO, (001) SrTiO$_3$
(STO), and (001) LaAlO$_3$ (LAO) single crystal substrates. The
calculated lattice mismatch values, defined as the percentage
difference between bulk FeSe and substrate in-plane lattice
parameters, are listed in Table I. The MgO substrate gives the
largest lattice mismatch, while the LAO substrate nearly matches the
lattice constants of bulk FeSe. The STO substrate has an
intermediate lattice mismatch. We studied 50 nm FeSe films on all
substrates, 100 nm on STO, and 200 nm on MgO and LAO. All films were
grown by using a pulsed laser deposition (PLD) technique. The
substrate temperature was kept at 380 $^0$C in 10$^{-6}$ torr vacuum
during the deposition. After the deposition, the samples were slowly
cooled down to room temperature in vacuum. The growth parameters
were studied carefully in order to get the best epitaxial growth of
tetragonal $\alpha$-FeSe films. The growth temperature is found to
be the most critical parameter. At high temperature ($\approx$ 800
$^0$C), no film was grown on the substrate. At a temperature between
450 $^0$C and 800 $^0$C, FeSe films tend to grow hexagonal and other
extra phases. Only those films grown at lower temperature ($\approx$
380 $^0$C) have the best tetragonal phase along with good mosaics.

A two-circle x-ray diffractometer (XRD) with monochromatic
CuK$\alpha$ ($\lambda$=1.540598 $\AA$) radiation was used to
determine the lattice constants of the FeSe films. The film
resistance was measured by using the 4-point technique. Temperatures
(5 K $\leq $ T $\leq$ 300 K) and magnetic fields (0 $\leq$ H $\leq$
9 T) were obtained using the Quantum Design PPMS system.


\begin{table}[htp]
\caption{Out-of-plane lattice constants (c) for FeSe films.
\label{modeltab}}
 \begin{tabular}{p{1.8cm} p{1.6cm}p{2.4cm}l}
 \hline
 Substrate\footnotemark & Lattice Mismatch  & Film Thickness & c-value \\
 \hline
 STO& 3.7 \% & 50  nm & 5.49 $\AA$\\
 MgO& 12 \% & 50  nm & 5.50 $\AA$ \\
 LAO& 0.64 \%  & 50  nm & 5.52 $\AA$\\
 STO& 3.7 \% & 100 nm & 5.51 $\AA$\\
 MgO& 12 \% & 200 nm & 5.52 $\AA$\\
 LAO& 0.64 \%  & 200 nm & 5.51 $\AA$\\
 \hline
 \end{tabular}
 \footnotetext[1]{ Bulk FeSe is tetragonal (a=b=3.765 $\AA$, c=5.518 $\AA$); STO is cubic (a=3.905 $\AA$), MgO is cubic (a=4.216 $\AA$); and LAO is
rhombohedral (a=3.789 $\AA$, c=13.11 $\AA$).}
 \end{table}
The XRD patterns for the three 50 nm films are shown in Fig.1 using
log scales. Panel (D) in Fig. 1 shows in detail the shift of the
(002) peak due to the effect of in-plane strain on the out-of-plane
lattice constant. The XRD patterns for the 100 nm and 200 nm FeSe
films are similar to those shown in Fig. 1 with relatively stronger
peaks. All of the tetragonal peaks were indexed using the P4/nmm
space group.  The FeSe films on STO show a pure tetragonal phase. No
obvious hexagonal FeSe (NiAs-type) were observed for all these
films. The extra peaks, which are due to second phases and the
substrate's background, are marked by stars (*). The sharp small
extra peaks are substrate K$_\beta$ peaks. The extra small broad
background for the films on LAO probably results from imperfections
of the LAO substrate. Overall, these results imply that the
$\alpha$-FeSe phase is dominant.

Out-of-plane lattice constants were calculated from the XRD data
using the (002) peaks (see Fig. 1), and are listed in Table I. In
our work, the FeSe films were found to be extremely sensitive to the
storage atmosphere. When exposed to air for a short period of time
($\approx$ hours), we observed the disappearance of the
superconducting transition. This is possibly due to oxidization or
other chemical reactions of the FeSe films. Epitaxial growth of the
FeSe film was checked by using a four-circle diffractometer (GADDS).
By comparing the out-of-plane lattice constants, we have calculated
the in-plane strain. It is reasonable to assume that the smaller the
out-of-plane lattice constants, the more tensile strain in the
plane.

The out-of-plane lattice constants show that the FeSe films have the
orientation with (00l) normal to the substrate surface. The c-values
for the 100 nm FeSe film on STO, 200 nm FeSe film on MgO and LAO are
5.51 $\AA$, 5.52 $\AA$, and 5.51 $\AA$, respectively. We consider
FeSe films with c = 5.51 $\AA$ and c = 5.52 $\AA$ as unstrained
since our best estimate of the combined systematic and random error
in the lattice constant to be about $\pm$ 0.01 $\AA$. Compared to
the bulk value of FeSe (c = 5.518 $\AA$), the three thicker FeSe
films are unstrained. This is because the 100 nm and 200 nm
thicknesses are above the relaxation critical thickness. The 50 nm
FeSe film on LAO are nearly unstrained due to the fairly good
lattice match. On the contrary, the 50 nm FeSe film on MgO is not
fully relaxed and under tensile strain. The tensile strain of the 50
nm FeSe film on STO is somewhat larger.

\begin{figure}[t]
\includegraphics[width=80mm]{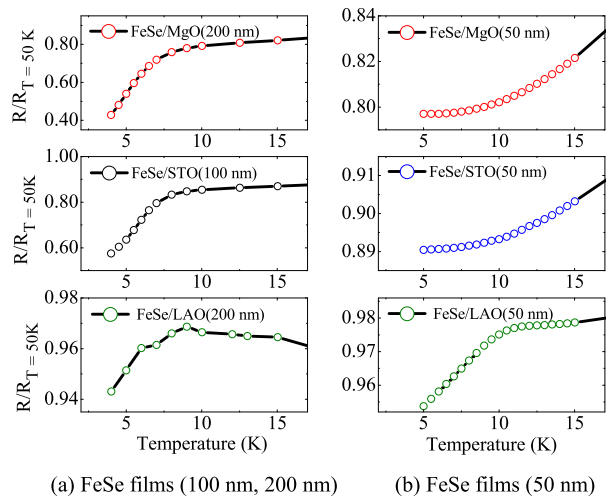}
\caption{\label{fig:fig1.eps} (color online) Resistance as a
function of temperature for: a) 100 nm and 200 nm and b) 50 nm FeSe
films on MgO, STO, and LAO substrates. }
\end{figure}


Figure 2 shows the resistance as a function of temperature for the
FeSe films on MgO, STO, and LAO substrates. An abrupt decrease of
the resistance at low temperature is observed for all thick (100 nm
and 200 nm) FeSe films, but only for one thin (50 nm) film, the film
on LAO. We attribute this drop to the onset of superconductivity.
This is because the film T$_{onset}$ values are approximately 8 K,
which coincide with the value for bulk polycrystalline
FeSe~\cite{Hsu01}. Furthermore, T$_{onset}$ exhibits a magnetic
field dependence; when the magnetic field increases, T$_{onset}$
decreases. Figure 3 shows the resistance as a function of
temperature with various external magnetic fields for 50 nm FeSe
film on LAO. The insert shows the external magnetic field as a
function of the superconducting onset temperature. The values for
T$_{onset}$ are chosen as the temperature where the resistance start
to decrease abruptly. The linearly extrapolated y-intercept value is
estimated to be 35 T, which is similar to the value for single
crystal FeSe~\cite{Patel01}, but about twice that for powder
samples~\cite{Hsu01}.

The resistance measurements described above demostrate that the
unstrained FeSe films show bulk-like properties with a T$_{onset}$
$\approx$ 8K. The 50 nm films on MgO and STO are under tensile
strain with no superconducting transitions down to 5 K. The strained
50 nm FeSe films are grown under the same condition as the 200 nm
FeSe films, and they are comparable in quality. The XRD patterns are
as good as the other films, the resistance is similar to the other
films, and most importantly, they are metallic down to 5K with no
upturn due to charge localization. This implies that the
superconductivity is not simply suppressed by having a poor quality
film that lacks a current path of the proper phase.  The tensile
strain itself suppresses the superconductivity.

\begin{figure}[t]
\begin{center}
\includegraphics[width=80mm]{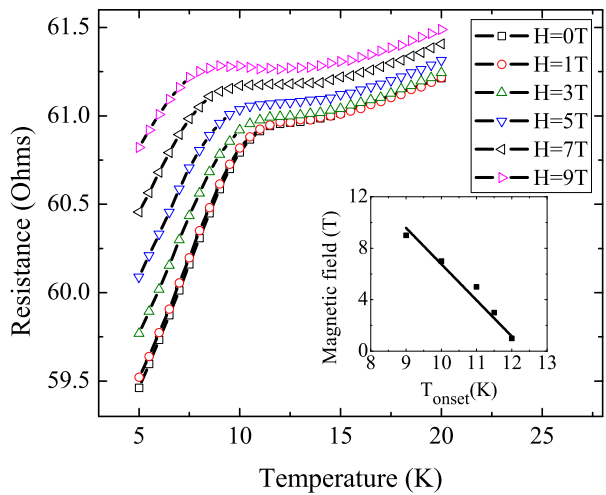}
\caption{\label{fig:fig1.eps} (color online) Resistance as a
function of temperature with various external magnetic fields for 50
nm FeSe film on LAO. The insert shows the external magnetic field as
a function of the superconducting onset temperature. }
\end{center}
\end{figure}



In summary, we grew 50 nm, 100 nm, and 200 nm tetragonal
$\alpha$-phase FeSe epitaxial films on various substrates by PLD.
The structure of the FeSe films depends strongly on the growth
conditions and the lattice mismatch with the substrates. All thick
(100 nm and 200 nm) films, and all films on LAO are mostly
unstrained, and behave like the bulk material. Thin (50 nm) FeSe
films on MgO and STO are under tensile strain. Evidence of
superconductivity has been observed in unstrained FeSe films, with
T$_{onset}$ close to that of bulk material. No sign of
superconductivity has been observed in FeSe films under tensile
strain down to 5 K. This implies that tensile strain suppresses
superconductivity.


This work is supported by the US-DOE through contract \#
DE-FG02-00ER45801.

\end{document}